# MEMORY AWARE HIGH-LEVEL SYNTHESIS FOR EMBEDDED SYSTEMS

Gwenole Corre, Eric Senn, Nathalie Julien, Eric Martin
*LESTER Laboratory, University of South-Brittany*
*BP92116, 56321 Lorient cedex, France*
*{gwenole.corre, nathalie.julien, eric.senn, eric.martin}@univ-ubs.fr*

**ABSTRACT**

We introduce a new approach to take into account the memory architecture and the memory mapping in the High-Level Synthesis of Real-Time embedded systems. We formalize the memory mapping as a set of constraints for the synthesis, and defined a Memory Constraint Graph and an accessibility criterion to be used in the scheduling step. We use a memory mapping file to include those memory constraints in our HLS tool GAUT. Our scheduling algorithm exhibits a relatively low complexity that permits to tackle complex designs in a reasonable time. Several experiments are performed to demonstrate the efficiency of our method, and to compare GAUT with an industrial behavioral synthesis tool. We finally show how to explore, with the help of GAUT, a wide range of solutions, and to reach a good tradeoff between time, power-consumption, and area.

## 1. INTRODUCTION

Behavioral synthesis, which is the process of generating automatically an RTL design from an algorithmic description, is an important research area in design automation. Many behavioral specifications, especially in digital signal and image processing, use arrays to represent, store and manipulate ever growing amounts of data. The ITRS roadmap (2001) indicates that, in 2011, 90 % of the SoC area will be dedicated to the memory. Applications are indeed becoming more and more complex, and memory will take a more and more important place in future signal processing systems. This place is vital, strategic, for memory now appears as a terrific bottleneck in real-time systems. Indeed, performances are highly dependent on the memory architecture (hierarchy, number of banks ) together with the way data are placed and transferred. Moreover, the design of the memory in a system has a very great impact on the power consumption, which is a so critical feature in embedded applications. Indeed, up to 80% of the global power consumption is due to the memory in current DSP applications Catthoor (1999). To tackle the complexity of memory design, we consider as essential to take into account memory accesses directly during the behavioral synthesis, assuming that a reasonable trade-off between the design time and the quality of the results is reached. In the context of HLS, several scheduling techniques actually include memory issues. Among them, most try to reduce the memory cost by estimating the needs in terms of number of registers for a given scheduling, but work only with scalars Gebotys (1997), Saied (1996). Some of them really schedule the memory accesses Passos (1995), Nicolau (1993). They include precise temporal models of those accesses, and try to improve performances without considering the possibility of simultaneous accesses which would ease the subsequent task of register and memory allocation. Works of Ellervee (2000) include the memory during HLS, but is dedicated to control intensive applications. In works of Seo (2001), a first scheduling (force directed) is performed on a Data Flow Graph (DFG); the memory accesses are then rescheduled after the selection and memory allocation to reduce the overall memory cost. The complexity of this scheduling algorithm, however, does not allow to target realistic applications in a reasonable time. Ly (1995) represents memory accesses as multi-cycle operations in a Control and Data Flow Graph (CDFG). Memory vertices are scheduled as operative vertices by considering conflicts among data accesses. This technique is used in some industrial

HLS tools that include memory mapping in their design flow (Monet, Behavioral Compiler) Knapp (1995). Memory accesses are regarded as Input/Output. The I/O behavior and number of control step are managed in function of the scheduling mode. In practice, the number of nodes in their input specifications must be limited, to obtain a realistic and satisfying architectural solution. This limitation is again mainly due to the complexity of the algorithms which are used for the scheduling. In this paper, we propose a new and simple technique to take into account the memory mapping in the architectural synthesis. Indeed, our aim is to produce a simple algorithm to achieve the synthesis of even complex designs in a reasonable time. In Section 2, we introduce our HLS tool: GAUT. We define its design flow and the architecture it produces. We focus on the definition of a memory mapping file that is used in the synthesis process. We introduce an original scheduling in the synthesis flow, to obtain an optimized RTL design. This scheduling technique is described in section 3 with the formalism to resolve scheduling under memory constraint. Optimizations are performed to reduce the needs in memory and processing units. Experimental results are discussed in section 4.

## 2. PRESENTATION OF THE HLS TOOL GAUT

GAUTis dedicated to real-time digital signal processing applications. It uses multifunctional (e.g. ALU) and pipeline operators. During the synthesis, a cost function, based on the circuit area and/or the power consumption, is optimized The resulting architecture finally supports the given timing constraint (the operating frequency). The real time constraint, provided by the designer as a data rate, represents the rates at which data is coming to the functional unit from an external source. This methodology not only reduces the design time, but also increases the reliability at every stage in the design flow. At last, hierarchical design is possible by including formerly synthesized operators. The design methodology follows the generic flow presented above with the following particularities: Compilation: parallelism extraction includes analysis, function in-lining, loop unrolling, and variable renaming (for single assignment). An internal SFG representation is used. Selection/Allocation: a greedy algorithm is applied. Scheduling/Assignment: achieved with a classical list scheduling algorithm. Binding: leaded by the branch and bound method

The synthesis process relies on a generic architecture based on the model of a digital signal processor core. It exploits spatial and temporal parallelisms to enhance computing. The generic architecture contains four functional entities: the processing unit, the memory unit, the communication unit and the control unit. The processing unit is synthesized first, for arithmetic treatments represent the bigger part of an algorithm. The tool insures that the real-time constraint is respected. The processing unit includes several cells. Each cell associates an operator (multiplier, ALU...), with multiplexors, demultiplexors, and registers. These cells communicate with a dedicated multi-bus. The number of cells, their constitution, as well as the number of busses, are optimized in order to minimize the cost function, while satisfying the timing constraint. The processing unit is pipelined when severe timing constraints are given. The memory unit deals with data storage. It is composed of registers, memory banks and address generators described as a Finite State Machine (FSM). During the processing unit design, whether the data is stored in register or memory is decided upon its lifetime (a threshold is settled by the designer; the lifetime of each variable is compared to this threshold). The communication unit is the interface with external circuits, such as another processor. The control unit controls the entire circuit. It is described as a Finite State Machine (FSM). More details about GAUT can be found in Martin (1993) or Julien (2003).

## 3. THE MEMORY

### 3.1. Memory aware synthesis

We introduce memory synthesis in the standard HLS design flow. The difference between the standard and the memory aware design flow is illustrated on Fig. 1. A Signal Flow Graph (SFG) is first generated from the algorithmic specification. In the new approach, this SFG is parsed and a memory table is created

(see Fig. 2). This memory table is then completed by the designer who can select the variable implementation (memory or register) and place the variable in the memory hierarchy (which bank). The resulting table is the memory mapping that will be used in the synthesis; memory accesses are now considered as a constraint for the scheduling. In the standard flow, the processing unit is synthesized without any knowledge on the memory mapping: the memory architecture is designed afterward and a lot of optimization opportunities are definitely lost.

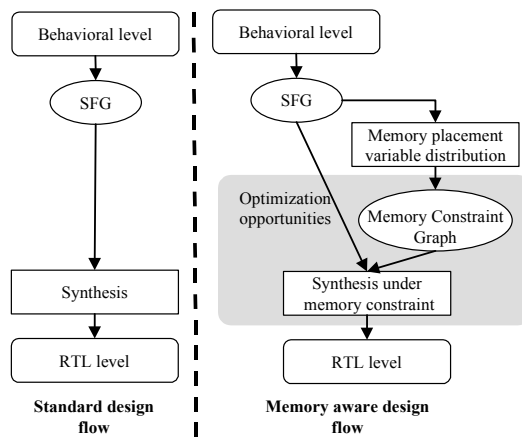

Figure 1 : standard and memory aware design flow

The memory mapping file contains information about every data structure in the algorithm (mainly arrays in DSP applications) and its allocation in memory (bank number and physical address). Scalars can also be defined. This memory table represents all data vertices extracted from a SFG. This data distribution can be static or dynamic. In the case of a static placement, the data stay at the same place during the whole execution. If the placement is dynamic, data can be transferred between different levels in the memory hierarchy. Thus, several data can share the same location in the circuit memory. The memory mapping file explicitly describes the data transfers to occur during the algorithm execution. Direct Memory Address (DMA) directives will be added to the code to achieve these transfers. The definition of the memory architecture will be performed in the first step of the overall design flow. To achieve this task, advanced compilers such as Rice HPF compiler, Illinois Polaris or Stanford SUIF could be used Panda (2001). Indeed, these compilers automatically perform data distribution across banks, determine which access goes to which bank, and then schedule to avoid bank conflicts. The Data Transfer and Storage Exploration (DTSE) method from IMEC and the associated tools (ATOMIUM, ADOPT) are also a good mean to determine a convenient data mapping Catthoor (2000).

| Name | Class | Implementation | Bank | Address | Initial Value |
|---|---|---|---|---|---|
| adapt | Variable | Register | -1 | -1 | 0 |
| deux_mu | Constant | Memory | 0 | 0 | 0 |
| h(0) | LoopBack | Memory | 0 | 1 | 0 |
| h(1) | LoopBack | Memory | 0 | 2 | 0 |
| h(2) | LoopBack | Memory | 0 | 3 | 0 |
| h(3) | LoopBack | Memory | 0 | 4 | 0 |
| x(0) | Variable | Memory | 0 | 9 | 0 |
| x(1) | Delay | Memory | 0 | 10 | 0 |
| x(2) | Delay | Memory | 0 | 11 | 0 |
| x(3) | Delay | Memory | 0 | 12 | 0 |

Figure 2 : Memory table

## 3.2. Signal Flow Graph

The input of our HLS tool is an algorithmic description specifies the circuit's functionality at the behavioral level, disregarding any potential implementation solutions. This initial description is compiled in order to obtain an intermediate representation: the Signal Flow Graph (SFG). Definition: A Signal Flow

Graph is a directed polar graph $SFG(V;E)$ where the set of vertices $V = \{v_0,...,v_n\}$ represents the operations, $v_0$ and $v_n$ are respectively the source vertex and the sink vertex. The set of edges $E = \{(vi; vj)\}$ represents the dependencies between the operations vertices. The Signal Flow Graph contains $|V| = n+1$ vertices. A vertex represents one of the following operations: arithmetic, logical, data or delay. The difference between a Signal Flow Graph and Data Flow Graph resides in the introduction of delay operators ($z^{-1}$). These operators are necessary to express the use of data whose value was computed in a preceding iteration of the algorithm. An edge $Ei; j = (vi; vj)$ represents a data dependence between operations $vi$ and $vj$ such as for any iteration of the SFG, operation $vi$ must start its execution before that of $vj$. For the data dependencies, the execution of $vj$ can start only after the completion of operation $vi$.

### 3.3. Memory Constraint Graph

As outlined in section 3.1, all data vertices are extracted from the SFG to construct the memory table. The designer can choose the data to be placed in memory and defines a memory mapping. For every memory in the memory table, we construct a weighted Memory Constraint Graph (MCG). It represents conflicts and scheduling possibilities between all nodes placed in this memory. The MCG is constructed from the SFG and the memory mapping file. It will be used during the scheduling step of the synthesis. Definition : a Memory Constraint Graph is a cyclic directed polar graph $MCG(V';E';W')$ where $V_0 = \{v'_0,...,v'_n\}$ is the set of data vertices placed in memory. A memory Constraint Graph contains $|V'| = n + 1$ vertices which represent the memory size, in term of memory elements. The set of edges $E_0 = \{(voi; voj)\}$ represents the precedence between the memory vertices, and W' is a function that represents the access delay between two data nodes. W' has only two possible values: Wseq (sequential) for an adjacent memory access in memory, or Wrand (randomize) for a non adjacent memory access. Weight depends on the data placement defined in the memory file. Fig. 3 shows a memory constraint graph for the LMS filter with two simple port memory banks. The input samples $x(i)$ are placed consecutively in one bank. The filter coefficients $h(i)$ are placed consecutively in one another bank (dotted edges represent edges where W = Wseq). The fastest sequence of accesses is found easily by following those dotted edges. In our example, the sequence {x0x1x2x3} is the fastest. Indeed, edges (x0x1) (x1x2) and (x2x3) are weighted with the minimal delay cost (dotted edges = Wseq); accesses to the memory are achieved in burst mode during this sequence. The fastest sequence in the MCG will be always selected as long as it respects the synthesis' constraints. Then, the scheduling will be directed to follow this sequence, and a maximum number of memory burst accesses will be provided. In the scheduling process, the MCG is also used to determine the accessibility criterion with the time of every memory access.

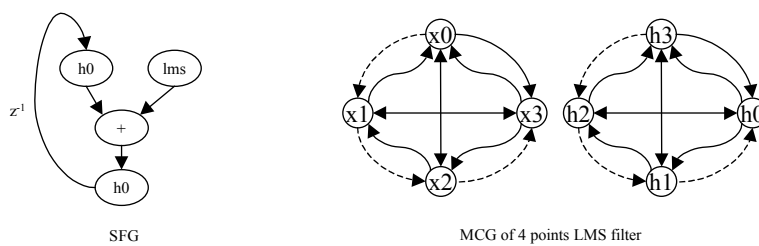

SFG                    MCG of 4 points LMS filter

Figure 3 : SFG and MCG examples

### 3.4. Scheduling under memory constraint

The classical list scheduling algorithm relies on heuristics in which ready operations (operations to be scheduled) are listed by priority order. In our tool, an early scheduling is performed. In this scheduling, the priority function depends on the mobility criterion. This mobility is computed, for each cycle, as the difference, in number of cycles, between the current cycle and the operation deadline. Whenever two ready operations need to access the same resource (this is a so called resource conflict), the operation with the lower mobility has the highest priority and is scheduled. The other is postponed. To perform a scheduling under memory constraint, we introduce fictive memory access operators and add an accessibility criterion based on the MCG. A memory has as much access operators as access ports. The memory is declared accessible if one of its fictive memory access operators is idle. Several operations can try to access the same memory in the same cycle; accessibility is used to determine which operations are really executable. Fictive memory access operators are represented by tokens on the MCG. There are as many tokens in the MCG as ports (R/W) in the memory. These tokens are used to compute the accessibility of the memory. The list of ready operations is still organized according to the mobility criterion, but all the operations that do not match the accessibility condition are removed from this list. To schedule an operation that involves an access to the memory, we check if the data is not in a busy memory bank. If a memory bank is not available, every operation that needs to access this memory will not be scheduled, no matter its priority level.

## 4. RESULTS

In this section, we first compare our tool GAUT with the industrial tool from Mentor Graphics: Monet. With the help of GAUT, we show the impact of data mapping on the final architecture, and how the data mapping can modify the circuit's size and power consumption. We also show how to use GAUT to find the better solution for a given timing constraint or for predetermined arithmetical resources. We finally use our approach to choose between different algorithms for a same application, and to analyze the complexity of each architectural solution.

### 4.1. GAUT vs Monet

Several syntheses were performed, both with GAUT and the industrial behavioral synthesis tool Monet. We chose the elliptic and the Kalman filters which are the biggest applications in the (HLSynth'92 benchmarks) , and two classical digital algorithms: a FIR filter and an echo cancellation algorithm, the LMS. Tab. 1, indicates the synthesis time in seconds and the architecture's latency in number of cycles (the same real-time constraint was given to the tools, the clock cycle is 10ns). Required hardware resources are also indicated: the number of registers (Reg), of multiplexers (Mux), demultiplexers (Demux), of glue logic elements (which are tri-states in GAUT), and the number of RAM and ROM memories. The two last columns give the number of read and write in those memories. Single port SRAM were used to store data. Syntheses were executed on SUN Blade 2000 workstations. Hardware resources are always lower in architectures synthesized with GAUT, although the same number of arithmetic operators is needed. The latency, which is the delay between the input of the first data and the first result on the output, is also lower with GAUT. It is necessary to distinguish three sorts of data in a signal processing application. First, there are the signals, which are the input and output flows of the applications. A mono-dimensional signal *x* is a vector of size *n*, if *n* values of *x* are needed to compute the result. Every cycle, a new value for *x* (*x*[*n* + 1]) is sampled on the input, and the oldest value of *x* (*x*[0]) is discarded. We called *x* an ageing, or maturing, vector or data. Second, there are the static coefficients, whose value is never changed. We chose to store these coefficients in ROM with GAUT, whereas they are wired with Monet. That explains why a ROM is needed with GAUT for the FIR filter, and not with Monet. Third, we consider the dynamic coefficients, whose value is changed during the execution of the algorithm, which is the case for an adaptative filtering like the LMS. Dynamic coefficients and ageing vectors are stored in RAM. In Monet, the new value of a signal is always written at the same address in memory, at the end of the vector in the case of a 1D signal for instance. That involves to shift every other values of the signal in the memory to free the place for the new value. This shifting necessitates *n* reads and *n* writes in the memory (and this is really time and power consuming). In GAUT, the new value is stored at the address of the oldest one in the vector. Only one write is needed. Obviously, the address generation is more difficult in this case, because the addresses of the samples called in the algorithm change from on cycle to the other. We have developed a new methodology to resolve the synthesis of these address generators. The advantage is a lower latency, since we avoid *n* reads and writes of the ageing vector, and a resulting lower power consumption. Indeed, the power consumption of a memory increases with the number of accesses. The synthesis time, together with the reduction of hardware resources and memory accesses, exhibit the efficiency of our scheduling technique. In fact, the difference between the synthesis time with GAUT and with a behavioral synthesizer like Monet increases with the complexity of the application. We have measured the synthesis times for the FIR and the LMS filters, with an increasing complexity. Tab. 2 presents the results for the FIR for 16, 32, 64, 128, 256, 512, and 1024 points. Tab 3 presents the results for the LMS filter for the same increasing complexities. It can be observed that, even if the difference between the synthesis time with GAUT and Monet is relatively small for small designs, it becomes enormous when the complexity increases. Indeed, it becomes hours, then days or weeks for the FIR 1024 and the LMS 512 and 1024. In fact, every memory access is a node to be schedule in Monet, and the scheduling algorithm has a strong complexity. The difference in latency is comparatively stable: the latency with Monet varies from about 2 to 3 times the latency with GAUT.

|  |  | Synth time | Lat(Nbcycle) | Reg | Mux | Demux | Tri | Glue | RAM | ROM | Nb read | Nb write |
|---|---|---|---|---|---|---|---|---|---|---|---|---|
| elliptic | Monet | 1s | 20 | 19 | 16 | 15 | – | 27 | – | – | – | – |
|  | Gaut | 1s | 20 | 12 | 6 | 9 | 24 | – | – | – | – | – |
| Kalman | Monet | 1s | 600 | 36 | 12 | 20 | – | 34 | – | – | – | – |
|  | Gaut | 1s | 60 | 14 | 11 | 10 | 29 | – | – | – | – | – |
| FIR 16 | Monet | 2s | 48 | 4 | 6 | 2 | – | 7 | 1 | – | 32 | 16 |
|  | Gaut | 1.4s | 19 | 4 | 2 | 1 | 1 | – | 1 | 1 | 32 | 1 |
| LMS 32 | Monet | 6s | 132 | 38 | 28 | 18 | – | 25 | 2 | – | 128 | 64 |
|  | Gaut | 1.4s | 100 | 19 | 3 | 3 | 23 | – | 2 | – | 128 | 33 |

Table 1 : GAUT vs MONET

| FIR | Tool | Nb cycle | Nb Read | Nb Write | Synth Time |
|---|---|---|---|---|---|
| 32 | Monet | 96 | 64 | 32 | 3 s |
|  | Gaut | 35 | 64 | 1 | 1.5 s |
| 64 | Monet | 192 | 128 | 64 | 7 s |
|  | Gaut | 67 | 128 | 1 | 1.7 s |
| 128 | Monet | 384 | 256 | 128 | 45 s |
|  | Gaut | 131 | 256 | 1 | 2 s |
| 256 | Monet | 768 | 512 | 256 | 20 m |
|  | Gaut | 259 | 512 | 1 | 3.8 |
| 512 | Monet | 1536 | 1024 | 512 | 7 h 17 |
|  | Gaut | 512 | 1024 | 1 | 4.9 |
| 1024 | Monet | 3072 | 2048 | 1024 | days |
|  | Gaut | 1027 | 2048 | 1 | 9 |

Table 2 : Syntheses of the FIR filter

| LMS | Tool | Nb cycle | Nb Read | Nb Write | Synth Time |
|---|---|---|---|---|---|
| 32 | Monet | 132 | 128 | 64 | 6 s |
|  | Gaut | 100 | 128 | 33 | 1.4 s |
| 64 | Monet | 260 | 256 | 128 | 14 s |
|  | Gaut | 196 | 256 | 65 | 1.9 s |
| 128 | Monet | 516 | 512 | 256 | 7 m 30 |
|  | Gaut | 388 | 512 | 129 | 2.6 s |
| 256 | Monet | 1028 | 1024 | 512 | 3 h 30 |
|  | Gaut | 772 | 1024 | 257 | 5.3 s |
| 512 | Monet | 2052 | 2048 | 1027 | days |
|  | Gaut | 1540 | 2048 | 513 | 9.6 s |
| 1024 | Monet | 4010 | 4096 | 2048 | weeks |
|  | Gaut | 3076 | 4096 | 1025 | 64 s |

Table 3 : Syntheses of LMS filter

## 4.2. Resources vs Performance Tradeoff

With the help of GAUT, it is easy to find the minimum number of operators and memory banks to satisfy the application's timing constraint. The results for a 32 points FFT are presented on Fig. 4. The application's data rate is given to the tool as the input data stream delay. When the data rate decreases, the number of simultaneous memory accesses, and so the number of memory banks, increases, as well as the number of operators. Given a number of banks, it is thus possible to find the minimum data rate, which is reached when the scheduling generates more simultaneous memory accesses than available memory access operators. In this case, the number of operating resources is also the biggest.

## 4.3. Memory exploration

Once found the required number of operators, one can try different numbers of memory banks as well as several data mappings, and evaluate their impact on the final application's performance and power consumption. In our example, for a data rate equals to 7.66s, we decide to allocate one multiplier and two ALU, and to share the memory in two banks. We then apply several memory mappings to the synthesis process, and observe the impact on the resulting circuit's power consumption. The circuit is a FPGA Xilinx Virtex XC400. Its consumption is computed with the Xilinx tool: XPower. One memory unit is generated for each memory mapping. The memory unit power consumption Pmu is provided in Tab. 4, together with the overall consumption Ptot, and the processing unit consumption Ppu. In the map2 16 mapping, the first sixteen real and imaginary FFT samples are mapped in the first bank, the remaining sixteen samples are in the second bank. In the map2 8, map2 4, and map2 2 mappings, samples are mapped respectively 8 by 8, 4 by 4, and 2 by 2, in the first and second memory banks. In the map2 1 mapping, even samples are in the first bank, odd samples in the second. Every memory unit invariably contains 2 RAM, and 4 FSM to drive the write and read accesses between the busses and the RAM. As a result, and because the number of memory accesses is also constant, there are very few variations on the memory unit power consumption (less than 2.5%, only due to small changes in I/O signals commutations and some logic blocs). Variations of the processing unit consumption are much more important, for they represent from 58% to 64.1% of the overall power consumption. The lowest overall consumption is obtained with the map2 4 mapping (35% lower than the map2 16), even if the memory unit consumption is slightly higher in this case (1.5%) than the lowest one.

|  | Pmu (mW) | Ppu (mW) | Ptot (mW) | ¢ (mW) | Ptot % |
|---|---|---|---|---|---|
| map2_16 | (2banks) | 12.71 | 34.35 | 47.06 | ... |
| map2_8 | (2banks) | 12.83 | 21.85 | 34.68 | -26 |
| map2_4 | (2banks) | 12.9 | 17.77 | 30.67 | -35 |
| map2_2 | (2banks) | 12.99 | 21.63 | 34.62 | -26 |
| map2_1 | (2banks) | 12.87 | 19.26 | 32.13 | -31 |

Table 4 : Power consumption

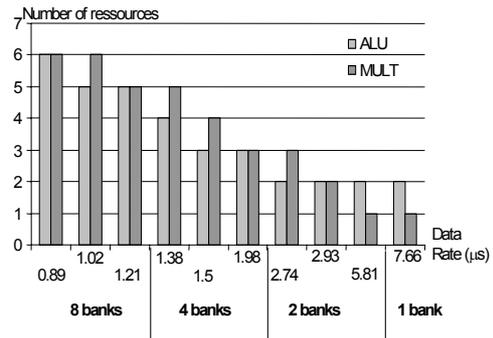

Figure 4 : Resources number vs data rate

## 4.4. Memory architecture for adaptive filtering

Two types of algorithm were developed for acoustic echo cancellation: recursive least squares and stochastic gradient. Algorithms of the first type are optimal in convergence, but have a very high complexity. On the contrary, algorithms of the second type exhibit a low complexity, but a rather poor convergence speed. In this section, we compare three algorithms of this second type: the LMS, the BLMS (with blocks of 4) and the GAL (with 10 cells). For two data rates, we carried out several synthesis to determine with the smallest processing unit, the best number of memory banks and the best data mapping for the memory unit. These optimal solutions are compared in Tab. 5. It is directly related to the number of logic blocks (CLBs) in the FPGA circuit that we targeted (again a Xilinx Virtex XC400). Tab. 5 indicates the memory unit's complexity. It is related to the number of memory banks, but also to the number of memory points and the number of accesses. As a result, the more interesting algorithm appears to be the LMS, for it always exhibits the lowest hardware complexity (the smallest number of CLBs, together with the smallest numbers of memory banks, memory points, and memory accesses). Obviously, the resulting circuit will also exhibit the smallest power consumption. Though the BLMS filter implies more memory accesses than the GAL, it also needs less memory points and less memory banks, and is therefore definitely less complex than the GAL. It will also consume less power.

| F (MHz) |  | Memory unit | | | Processing unit |
|---|---|---|---|---|---|
|  |  | locations number | accesses number | banks number | CLBs number |
| 1 | LMS | 65 | 160 | 3 | 504 |
|  | BLMS | 135 | 842 | 4 | 640 |
|  | GAL | 217 | 575 | 8 | 1168 |
| 2 | LMS | 90 | 210 | 4 | 904 |
|  | BLMS | 132 | 694 | 6 | 1192 |
|  | GAL | 154 | 422 | 10 | 1736 |

Table 5 : Complexity of memory and processing units

## 5. CONCLUSION

In this paper, we present a new strategy to take into account the memory architecture and the memory mapping in High- Level Synthesis. We define the memory mapping constraint and include it in the synthesis design flow. We introduce Memory Constraint Graphs, and an accessibility criterion to enhance the scheduling algorithm. Our method was included in GAUT, the HLS tool developed in the LESTER Laboratory. Several experiments were made, to explore the efficiency of our approach. The comparison with an industrial behavioral synthesis tool exhibits several advantages for GAUT. It appears firstly that GAUT uses less hardware resources, and reduces the count of memory accesses, which lead to a lower latency and a lower power consumption. Secondly, GAUT is able to tackle complex designs, and to perform the synthesis in a reasonable time. Memory aware synthesis and GAUT appear very efficient for exploring the design space and for balancing optimizations between the processing unit and the memory unit. It permits to determine the best memory architecture, i.e. the best number of memory banks, as well as the best memory mapping, to meet the application constraints, and to finally reach a reasonable tradeoff between time, power consumption, and area. Future works aim to improve the scheduling step with an anticipated read model for the data, which should allow to speedup the processing unit. Efforts will be made on the automatic determination of the number of memory bank from the application data rate. Furthermore, the data stability criterion will be included to reduce the activity on the operator's inputs and to further reduce the power consumption.